\documentclass[acmsmall,screen,nonacm]{acmart}

\usepackage[ruled,vlined,linesnumbered]{algorithm2e}
\SetKwProg{Fn}{Function}{:}{}
\SetKwFunction{FProx}{ProximityProfile}
\usepackage{graphicx}
\usepackage{textcomp}
\usepackage{subfig}  
\usepackage{tikz}
\usepackage{multirow}
\usepackage{colortbl}
\usepackage{comment}
\usepackage{placeins}
\usepackage{cleveref}
\usepackage{xcolor}
\usepackage{tcolorbox}
\usepackage[normalem]{ulem}

\tcbuselibrary{skins,breakable}

\definecolor{winered}{rgb}{0.5, 0.0, 0.0}
\definecolor{navyblue}{rgb}{0.0, 0.0, 0.5}
\AtBeginDocument{%
  \hypersetup{linkcolor=winered,citecolor=winered,urlcolor=navyblue,
    anchorcolor=winered,filecolor=winered}}

\newcommand{\rz}[1]{\textcolor{red}{[RZ: #1]}}
\newcommand{\zy}[1]{\textcolor{purple}{[ZY: #1]}}
\newcommand{\gz}[1]{\textcolor{cyan}{[GZ: #1]}}
\newcommand{\note}[1]{\textcolor{winered}{\textbf{#1}}}
\definecolor{unsureorange}{HTML}{B45309}
\definecolor{blablagreen}{HTML}{15803D}
\newcommand{\unsure}[1]{{#1}}
\newcommand{\blabla}[1]{{#1}}

\newcommand{\prox}{\operatorname{\text{\normalfont\scshape prox}}}

\NewDocumentCommand{\delbyzy}{O{ZY} m}{%
  \textcolor{purple}{\sout{#2}}\textsuperscript{\textcolor{purple}{[#1]}}%
}

\newcommand{\sector}[1]{\makebox[1.25em][l]{\ensuremath{#1}:}}
\newcommand{\slot}[1]{\makebox[1.6em][c]{\ensuremath{#1}}}

\Crefname{theorem}{Theorem}{Theorems}
\newtheorem{definition}{Definition}\Crefname{definition}{Definition}{Definitions}
\Crefname{lemma}{Lemma}{Lemmas}
\Crefname{corollary}{Corollary}{Corollaries}
\Crefname{proposition}{Proposition}{Propositions}
\Crefname{remark}{Remark}{Remarks}

\definecolor{ElegantGrayBack}{RGB}{242, 242, 242} 
\definecolor{ElegantGrayFrame}{RGB}{105, 105, 105} 

\newtcolorbox{takeaways}[1][]{
    colback=ElegantGrayBack, 
    colframe=ElegantGrayFrame, 
    fonttitle={\small\bfseries},
    title={Takeaways},       
    arc=1.5mm,                 
    boxrule=0.8pt,           
    left=5pt, right=5pt,     
    top=3pt, bottom=3pt,     
    enhanced,                
    breakable,               
    #1                       
}

\begin{document}

\title{Proxima: Optimizing Syndrome Extraction Circuit Scheduling for Practical Quantum LDPC Codes}

\author{Ruiyun Zhang}
\authornote{Both authors contributed equally to this work.}
\affiliation{%
  \institution{The Hong Kong University of Science and Technology}
  \country{Hong Kong}}
\author{Guangqi Zhao}
\authornotemark[1]
\affiliation{%
  \institution{MatriQ Computing}
  \country{China}}
\author{Zhaohui Yang}
\authornote{Corresponding author: \href{mailto:zhaohui@ucsb.edu}{zhaohui@ucsb.edu}.}
\affiliation{%
  \institution{The Hong Kong University of Science and Technology}
  \country{Hong Kong}}
\author{Yingjian Liu}
\affiliation{%
  \institution{Leiden University}
  \city{Leiden}
  \country{Netherlands}}
\author{Chenghong Zhu}
\affiliation{%
  \institution{QudeLeap Research}
  \country{China}}
\author{Anbang Wu}
\affiliation{%
  \institution{Shanghai Jiao Tong University}
  \city{Shanghai}
  \country{China}}
\author{Jun Xia}
\affiliation{%
  \institution{The Hong Kong University of Science and Technology}
  \country{Hong Kong}}
\author{Yuan Xie}
\affiliation{%
  \institution{The Hong Kong University of Science and Technology}
  \country{Hong Kong}}

\renewcommand{\shortauthors}{Ruiyun Zhang et al.}

\authorsaddresses{}

\begin{abstract}

The fault-tolerant performance of quantum error correction codes relies critically on the scheduling of parity checks in syndrome extraction circuits (SECs). For general CSS-type quantum low-density parity-check (qLDPC) codes, the vast scheduling space and intricate error-propagation patterns make it computationally prohibitive to directly optimize the logical error rate (LER) or the exact circuit-level distance ($d_{\mathrm{circ}}$). To address this challenge, we introduce Proxima, a highly effective automated SEC scheduling framework guided by a distance-related metric termed \emph{proximity}. Defined directly on the detector error model (DEM), proximity quantifies the minimum number of additional faults required to complete a seed fault to an undetectable logical failure, naturally capturing arbitrary circuit-level fault mechanisms and bounding $d_{\mathrm{circ}}$ from above. Proxima synthesizes a pool of valid schedules at a target depth and ranks them by evaluating proximity within small, graph-local DEM fault subsets, progressively enlarging the subset only to break ties among top-ranked candidates. Across a broad spectrum of qLDPC code families, Proxima consistently attains low circuit depths and high circuit-level distance bounds, reducing LER by 1.4--2.6$\times$ over the strongest automated baseline and up to $40\times$ over others while generalizing robustly across noise models and decoders. Furthermore, by exploiting translational symmetry via a compact monomial representation, Proxima discovers depth-8 uniform schedules for bivariate bicycle codes that surpass even IBM's depth-7 reference schedule [Nature, 2024] by 2.1--5.1$\times$, establishing the DEM-level proximity as a practical objective for fault-tolerant compilation.

\end{abstract}


\maketitle

\section{Introduction}
\label{sec:introduction}


Quantum error correction (QEC) is essential for protecting fragile quantum information against noise and operational imperfections. In stabilizer codes, logical information is encoded into a codespace defined as the $+1$ co-eigenspace of a set of mutually commuting Pauli operators, termed stabilizers~\cite{Gottesman1997Thesis}. Pauli errors move quantum states out of the codespace by anticommuting with stabilizers. Identifying and diagnosing these errors relies on repeatedly measuring these stabilizers, a procedure known as syndrome extraction~\cite{nielsen2010quantum}. In practice, syndrome extraction is implemented through a circuit, where each stabilizer generator is assigned an ancilla qubit that interacts sequentially with the data qubits in its support through two-qubit entangling gates, typically CNOTs. Subsequent measurement of the ancilla yields the syndrome, providing necessary error diagnostics while preserving the encoded logical state.


While the fault-tolerant performance of a QEC code, commonly characterized by the logical error rate (LER), depends on its code distance~\cite{Dennis2002Topological,fowler_surface_2012} and decoding strategies~\cite{battistel2023real}, it is also governed by the design of \emph{syndrome extraction circuits} (SECs)~\cite{zhu2026quantum,Derks2025designingfault}. Even at a fixed circuit depth, there remains substantial freedom in scheduling the sequence of two-qubit entangling gates between data and ancilla qubits.  
An unfavorable schedule can induce error propagation that drives the circuit-level distance below the code distance. 

SEC scheduling is therefore a first-order design consideration, especially for general quantum low-density parity-check (qLDPC) codes~\cite{Breuckmann2021ldpc}, whose high-weight and nonlocal checks admit an exponentially larger search space than regular surface codes~\cite{fowler_surface_2012}. However, neither LER nor circuit-level distance can serve directly as a practical optimization objective. Evaluating LER requires compiling a \emph{detector error model} (DEM)~\cite{gidney2021stim,Higgott2025sparseblossom,Derks2025designingfault}, sampling millions of circuit faults, and running decoders, while computing the exact circuit-level distance demands an NP-hard combinatorial search over the full DEM graph. Practical SEC schedule optimization thus hinges on surrogate metrics that are computationally tractable yet highly predictive of logical performance.

\begin{table}[t]
  \centering
  \caption{Taxonomy of representative automated SEC scheduling methods.}
  \label{tab:taxonomy}
\footnotesize
\renewcommand{\arraystretch}{1.25}
\setlength{\tabcolsep}{4pt}
\newcolumntype{Y}[1]{>{\raggedright\arraybackslash}p{\dimexpr(\linewidth-12\tabcolsep)*#1/1000\relax}}
\definecolor{proximarow}{HTML}{E8EEF6}
\begin{tabular}{Y{146}Y{184}Y{184}Y{176}Y{131}Y{179}}
  \toprule
  \textbf{Method} & \textbf{Schedule generation} & \textbf{Objective} & \textbf{Faults seen by the objective} & \textbf{$X$/$Z$ ordering} & \textbf{Cost to score one schedule} \\
  \midrule
  ASC~\cite{zhang_optimal_2026} & SMT search over valid schedules & Circuit depth & None & Interleaved & Negligible \\
  \addlinespace[1pt]
  LRC~\cite{strikis_high-performance_2026} & Left/right partition and edge coloring; color permutations & Residual distance and its profile & Single ancilla faults completed by data faults (code capacity) & Non-interleaved & Code-capacity distances, precomputed per residual error \\
  \addlinespace[1pt]
  AlphaS.~\cite{liu_alphasyndrome_2026} & Monte Carlo tree search & Logical error rate, decoder in the loop & Circuit-level noise, sampled over one noisy round & Alternating-$X$/$Z$\textsuperscript{$*$} & Stim sampling and decoding at every rollout \\
  \addlinespace[1pt]
  PropHunt~\cite{viszlai_prophunt_2026} & Iterative CNOT reordering, starting from a coloration circuit & Syndrome ambiguity in randomly grown DEM subgraphs & All circuit fault mechanisms (DEM) & Interleaved & MaxSAT on 500 sampled subgraphs per iteration \\
  \addlinespace[1pt]
  \rowcolor{proximarow}
  \textbf{Proxima} & Valid schedules via SMT at a target depth, ranked by proximity & Proximity as an upper bound on circuit-level distance & All circuit fault mechanisms (DEM) & Interleaved & SMT on local column subsets of the DEM \\
  \bottomrule
\end{tabular}

  \begin{flushleft}
    \footnotesize
    $^{*}$Alternating-$X$/$Z$ is an extreme case of non-interleaved scheduling, where all $X$-check CNOTs are scheduled before all $Z$-check CNOTs or vice versa.
  \end{flushleft}
\end{table}

Early SEC designs are predominantly derived through code-specific heuristics~\cite{Low_distance,bravyi_high-threshold_2024,Constant_overhead,Lee2025colorcodedecoder}. Representative examples include the hand-crafted N--Z CNOT ordering for the rotated surface code, which keeps hook errors from aligning with logical operators~\cite{Low_distance}, and the depth-optimal interleaved parity-check schedule for bivariate bicycle (BB) codes, found by searching over gate orderings that satisfy the commutation constraints and maintain a good circuit-level distance~\cite{bravyi_high-threshold_2024}. However, these bespoke strategies neither guarantee optimality nor generalize to other code families. Recently, automated SEC scheduling methods have been proposed to close this gap. For instance, Auto-Stabilizer-Check (ASC)~\cite{zhang_optimal_2026} minimizes circuit depth by casting parity-check ordering as a satisfiability modulo theories (SMT) problem; \citet{tan_syndrome_2026} exploit the algebraic structure of quasi-abelian lifted product codes, including hypergraph product and bicycle codes, to construct schedules within one layer of the depth lower bound; PropHunt~\cite{viszlai_prophunt_2026} minimizes syndrome ambiguity, i.e., the extent to which distinct faults share a syndrome yet require different corrections; AlphaSyndrome~\cite{liu_alphasyndrome_2026} navigates the schedule space via Monte Carlo tree search (MCTS) with a decoder in the loop, so every search step pays for simulation and decoding and the resulting schedule is tailored to that decoder; and the left--right circuit (LRC) construction~\cite{strikis_high-performance_2026} partitions the data qubits into left and right sets, staggers the $X$- and $Z$-checks without interleaving their gates, and ranks circuits by the code-capacity residual distance, which counts an ancilla fault together with the fewest data faults that complete its hook error into a logical operator.


Together, these efforts highlight the diverse strategies employed to optimize SEC schedules, ranging from heuristic constructions to automated search guided by surrogate metrics. Minimizing circuit depth reduces idle errors and shortens logical clock cycles but ignores fault propagation entirely~\cite{zhang_optimal_2026,tan_syndrome_2026,durazo_rocha_local_2026}, while metrics like syndrome ambiguity~\cite{viszlai_prophunt_2026} and residual distance~\cite{strikis_high-performance_2026} capture only specific aspects of error propagation, leaving room for more comprehensive and predictive surrogate objectives. Conversely, search-based methods that estimate the LER of every candidate by simulation and decoding~\cite{liu_alphasyndrome_2026,ye_reinforcement_2026} incur prohibitive computational overhead and do not exploit the structure of QEC codes. This exposes a fundamental gap between scalability and physical reality: effective SEC optimization requires elevating full circuit-level error propagation into an analytical, structure-aware compiler objective, avoiding both the blind spots of partial heuristics and the prohibitive overhead of in-loop simulation.

To bridge this gap, we present Proxima, an automated SEC scheduling framework for CSS-type qLDPC codes. At the core of Proxima is \emph{proximity}, an analytical surrogate metric defined directly on the DEM that quantifies how close individual circuit faults are to forming an undetectable logical error. By expanding local graph neighborhoods of faults, Proxima evaluates proximity exactly inside each neighborhood at a fraction of the computational cost of full distance verification. As summarized in \Cref{tab:taxonomy}, Proxima is the only automated method that scores every candidate with a deterministic objective covering all circuit fault mechanisms while keeping interleaved $X$/$Z$ schedules in its search space.
Our key contributions are as follows:
\begin{itemize}
    \item We propose {proximity}, a distance-related metric on the DEM that covers all circuit fault mechanisms and bounds the circuit-level distance from above.
    \item We present Proxima, a highly effective SEC optimization framework that generates valid SEC schedules at a target depth and ranks them by proximity over progressively enlarged fault neighborhoods.
    \item We specialize Proxima to BB codes through a compact monomial representation, which leverages their translational symmetry and maps gate orderings to permutations of monomials. Proxima identifies a uniform depth-8 schedule that reduces the LER by 52--81\% compared to IBM's depth-7 reference schedule~\cite{bravyi_high-threshold_2024}.
    \item Across a variety of qLDPC codes, {Proxima} consistently outperforms current SOTA automated schedulers (ASC, LRC, AlphaSyndrome, and PropHunt) and attains low circuit depths and high circuit-level distance bounds, reducing LER by 1.4--2.6$\times$ over the strongest baseline and maintaining its advantage across alternative noise models and decoders.    
\end{itemize}

\section{Background}
\label{sec:background}

\subsection{Quantum Stabilizer Codes}


Stabilizer codes~\cite{Gottesman1997Thesis} form a structured family of quantum error-correcting codes. An $[[n,k,d]]$ stabilizer code encodes $k$ logical qubits into $n$ physical qubits and has distance $d$, defined as the minimum weight of a nontrivial logical Pauli operator. The codespace is the simultaneous $+1$ eigenspace of an Abelian subgroup $\mathcal{S}$ of the $n$-qubit Pauli group, with $-I\notin\mathcal{S}$. The elements of $\mathcal{S}$ act as parity checks: any Pauli error that anticommutes with at least one stabilizer changes the corresponding syndrome and moves the state out of the codespace. Logical Pauli operators are elements of the normalizer $N(\mathcal{S})$ that are not themselves stabilizers, i.e., elements of $N(\mathcal{S})\setminus\mathcal{S}$. The code distance is therefore $d=\min_{P\in N(\mathcal{S})\setminus\mathcal{S}} \operatorname{wt}(P)$.

\begin{figure}[tbp]
  \centering
  \subfloat[BB code structure example.\label{fig:bb_code_structure}]{%
    \includegraphics[height=0.45\columnwidth]{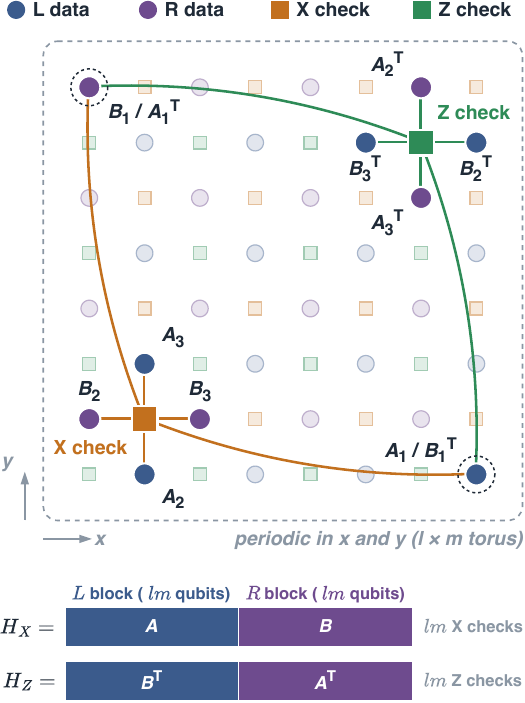}}%
  \qquad
  \subfloat[BB code SEC schedule example.\label{fig:bravyi_ancilla_view_schedule}]{%
    \includegraphics[height=0.45\columnwidth]{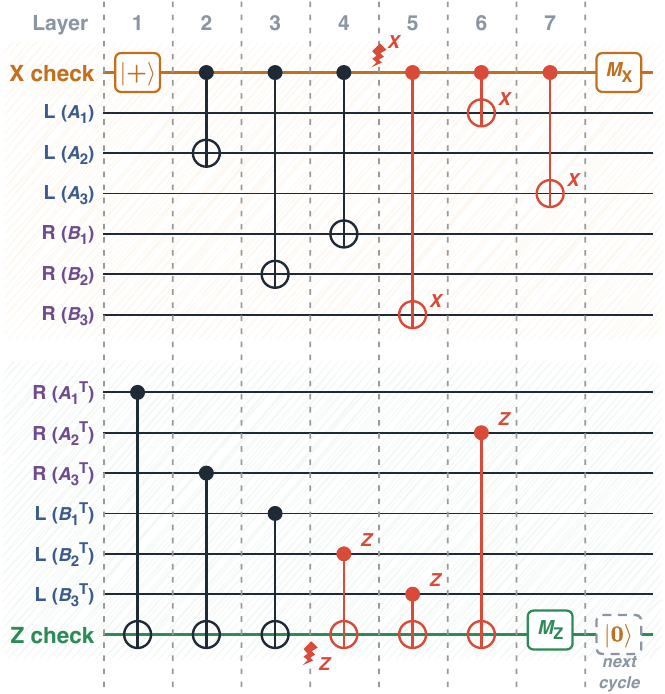}}
  \caption{BB code structure and an example SEC schedule. (a) $[[72,12,6]]$ BB code example, with $A=x^3+y+y^2$ and $B=y^3+x+x^2$. (b) A reference SEC schedule in the ancilla-centric view. Each panel follows one representative check and the six data qubits in its support, with the $X$ check in the order $A_2 B_2B_1B_3A_1A_3$ and the $Z$ check in the order $A_1^{\mathsf T}A_3^{\mathsf T}B_1^{\mathsf T}B_2^{\mathsf T}B_3^{\mathsf T}A_2^{\mathsf T}$; translational symmetry allows all checks of the same type to follow the same sequence. To illustrate hook errors, an $X$ ($Z$) fault on the $X$-check ($Z$-check) ancilla right after layer 4 (3) propagates through the remaining CNOTs and leaves $X$ ($Z$) errors on three data qubits.}
  \label{fig:bb-code-example}
\end{figure}

Calderbank--Shor--Steane (CSS) codes~\cite{calderbank1996good,steane1996error} form a special class of stabilizer codes in which the stabilizer generators can be partitioned into purely $X$-type and $Z$-type checks. These two sets of checks are naturally represented by binary parity-check matrices $H_X$ and $H_Z$, with the commutation condition reducing to $H_X H_Z^{\mathsf T}=0 \pmod 2$. This structure cleanly separates $X$- and $Z$-type error correction and logical operators, thus simplifying both code construction and syndrome extraction circuit design. CSS codes include many common code families, such as toric codes, rotated surface codes, color codes, and most qLDPC codes~\cite{Dennis2002Topological,bombin2006topological,fowler_surface_2012,Breuckmann2021ldpc,bravyi_high-threshold_2024}.

qLDPC codes are especially characterized by their sparsity structure~\cite{Breuckmann2021ldpc}. Both the weights of stabilizer checks and the number of checks acting on each data qubit remain bounded independently of the code size $n$. This locality in the Tanner-graph sense facilitates efficient syndrome extraction and scalable implementation, since bounded-weight checks require only a constant number of ancilla--data interactions per stabilizer. More crucially, qLDPC codes promise fault tolerance with constant qubit overhead~\cite{gottesman2014faulttolerant} at thresholds on par with that of the surface code~\cite{bravyi_high-threshold_2024}, making them strong candidates for future fault-tolerant quantum computing.

Among CSS-type qLDPC code families, bivariate bicycle (BB) codes~\cite{bravyi_high-threshold_2024} are a practically relevant class defined on an $\ell\times m$ torus. Let $S_\ell$ denote the $\ell\times\ell$ cyclic shift matrix, and let $x=S_\ell\otimes I_m$ and $y=I_\ell\otimes S_m$, which commute with each other. A BB code is specified by two matrices $A$ and $B$, each a sum of three distinct monomials in $x$ and $y$, which define its parity-check matrices as
\begin{equation}
  A=A_1+A_2+A_3,\quad B=B_1+B_2+B_3,\quad H_X=[A\mid B],\quad H_Z=[B^{\mathsf T}\mid A^{\mathsf T}].
  \label{eq:bb-defining-polynomials}
\end{equation}
Since $A$ and $B$ commute, $H_XH_Z^{\mathsf T}=AB+BA=0 \pmod 2$ and the CSS condition holds. The code has $n=2\ell m$ data qubits, split into a left ($L$) and a right ($R$) block of $\ell m$ qubits each, and $\ell m$ checks of each type. Every $X$ check acts on three $L$-qubits via $A_1,A_2,A_3$ and three $R$-qubits via $B_1,B_2,B_3$, while every $Z$ check acts on three $L$-qubits via $B_i^{\mathsf T}$ and three $R$-qubits via $A_i^{\mathsf T}$. All checks therefore have weight 6, each data qubit belongs to three checks of each type, and checks of the same type are translations of one another on the torus. \Cref{fig:bb_code_structure} illustrates this structure on the $[[72,12,6]]$ code with $\ell=m=6$.

\subsection{Syndrome Extraction Circuits}
\label{sec:background:sec}


\begin{figure}[tbp]
  \centering
  \captionsetup[subfloat]{farskip=0pt}
  \includegraphics[width=0.7\linewidth,trim={0 151bp 0 0},clip]{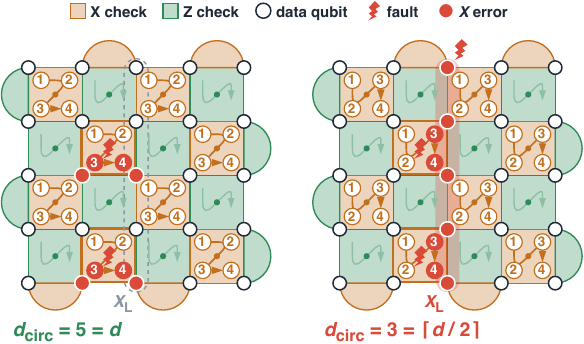}\par
  \subfloat[Z-shaped $X$-check order.\label{fig:hook-z-order}]{%
    \includegraphics[width=\dimexpr 0.7\linewidth*131/280\relax,trim={0 0 149bp 14bp},clip]{figures/hook_effective_distance.pdf}}%
  \hspace{\dimexpr 0.7\linewidth*18/280\relax}%
  \subfloat[N-shaped $X$-check order.\label{fig:hook-n-order}]{%
    \includegraphics[width=\dimexpr 0.7\linewidth*131/280\relax,trim={149bp 0 0 14bp},clip]{figures/hook_effective_distance.pdf}}
  \caption{Hook errors and effective distance on the $d=5$ rotated surface code. Orange (green) plaquettes are $X$ ($Z$) checks, half-disks are weight-2 boundary checks, and badges give the CNOT order of the $X$ checks; the $Z$ checks use the N-shaped order in both panels. A fault on an $X$-check ancilla after its second CNOT propagates $X$ onto the targets of CNOTs 3 and 4. (a) With the Z-shaped order the two errors of each hook lie across the logical operator $X_L$, no logical string is shortened, and the circuit-level distance stays at 5. (b) With the N-shaped order they lie along $X_L$: two hooks in consecutive rounds plus one data fault complete $X_L$ without triggering any detector, so three faults suffice and $d_{\mathrm{circ}}=3=\lceil d/2\rceil$. Both schedules are valid syndrome-extraction circuits; only the gate order differs.}
  \label{fig:hook-error-impact}
\end{figure}

In the circuit model of quantum computation~\cite{nielsen2010quantum}, fault-tolerant memory with stabilizer codes is maintained through repeated rounds of syndrome extraction using ancillary qubits. The corresponding SEC couples an ancilla to the data qubits in the support of each stabilizer generator, typically via two-qubit entangling gates such as CNOT or CZ gates. Subsequent projective measurements of the ancillas yield the error syndrome, from which a classical decoder infers the most likely underlying error configuration to guide recovery.

Typically, for general qLDPC codes, circuit depth is an important design criterion, indicating the duration of each syndrome extraction cycle. Even at a fixed depth, however, there remains substantial freedom in ordering the ancilla--data interactions. Subject to the commutation constraints among $X$- and $Z$-type checks~\cite{zhang_optimal_2026}, there exist many valid gate ordering schemes, referred to as different \emph{schedules}. The scheduling choice is critical because faults on an ancilla can propagate through subsequent gates to multiple data qubits, producing correlated errors known as \emph{hook errors}~\cite{Dennis2002Topological,Yoder2017surfacecodetwist}. When these propagated errors align with logical operators, a logical failure can be caused by fewer circuit faults in total. This dynamic degradation is quantified by the \emph{circuit-level distance} $d_{\mathrm{circ}}$, defined as the minimum number of circuit faults that produce an undetectable, logically nontrivial operator. An unfavorable schedule can drive $d_{\mathrm{circ}}$ well below the nominal code distance $d$. For example, on the $d=5$ rotated surface code in \Cref{fig:hook-error-impact}, the Z-shaped order of the $X$ checks preserves $d_{\mathrm{circ}}=5$, whereas the N-shaped order aligns the hook errors with $X_L$ and reduces it to $3$~\cite{fowler_surface_2012}. \Cref{fig:bravyi_ancilla_view_schedule} further illustrates a depth-7 reference SEC schedule of \citet{bravyi_high-threshold_2024} for BB codes, in which a single ancilla fault after the third CNOT of a check incurs a hook error on three data qubits.

\subsection{Detector Error Model}
\label{sec:background:dem} 

Under circuit-level noise, gates, measurements, and idling can all introduce faults during syndrome extraction, making a single measurement round insufficient to distinguish between operation and measurement errors. The \emph{detector error model} (DEM)~\cite{Higgott2025sparseblossom,Derks2025designingfault} provides a convenient representation of faults in such repeated Clifford circuits. A \emph{detector} is defined as a parity of measurement outcomes that is deterministic in the absence of faults. QEC typically repeats syndrome extraction for $d$ (code distance) rounds and jointly decodes the resulting space-time syndrome data~\cite{gottesman2026surviving}.

\begin{figure}[tbp]
  \centering
  \subfloat[Memory circuit with faults.\label{fig:dem-sec}]{%
    \includegraphics[height=0.31\linewidth]{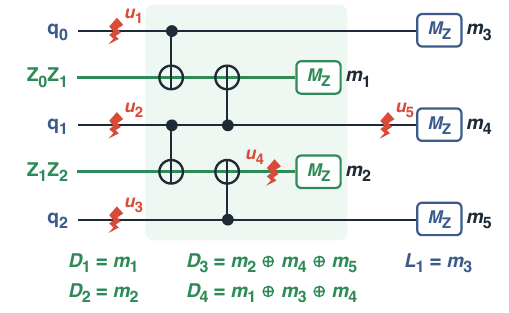}}%
  \hfill
  \subfloat[Resulting DEM graph.\label{fig:dem-graph}]{%
    \includegraphics[height=0.31\linewidth]{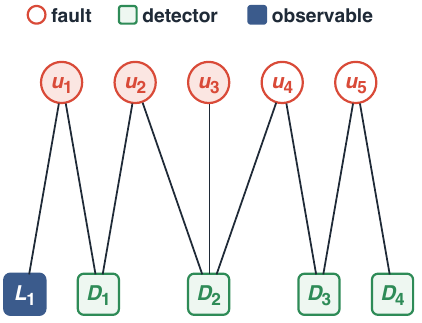}}
  \caption{DEM of a distance-3 repetition-code memory with one round of syndrome extraction (green background) and a final data readout. (a) Each fault $u_i$ flips some of the outcomes $m_1,\ldots,m_5$, and the detectors $D_j$ and the observable $L_1$ are parities of these outcomes. (b) The shaded faults $u_1,u_2,u_3$ flip $L_1$ but no detector, and no smaller set of circuit faults does so, hence $d_{\mathrm{circ}}=3$.}
  \label{fig:dem-example}
\end{figure}

Following the convention from Stim~\cite{gidney2021stim}, an SEC contains $N_f$ elementary fault mechanisms $\mathcal{U} = \{u_1, u_2, \ldots, u_{N_f}\}$, each flipping a subset of raw measurement outcomes. Let $N_D$ and $N_L$ denote the numbers of detectors and observables, respectively. By composing the fault-to-measurement flips with the measurement-to-detector and measurement-to-observable parity maps over $\mathbb{F}_2$, the DEM is compactly represented by a \emph{detector error matrix} $H\in\mathbb{F}_2^{N_D\times N_f}$, an \emph{observable matrix} $L\in\mathbb{F}_2^{N_L\times N_f}$, and the prior probability for each fault mechanism~\cite{Derks2025designingfault}. Column $i$ of $H$ ($L$) marks the detectors (observables) flipped by $u_i$. The DEM can be represented as a Tanner graph, dubbed the \emph{DEM graph} and depicted in \Cref{fig:dem-example}, in which fault node $u_i$ is adjacent to detector node $D_j$ if $H_{j,i}=1$ and to observable node $L_k$ if $L_{k,i}=1$. A fault set, encoded by $x\in\mathbb{F}_2^{N_f}$, is undetectable if $Hx=0$ and logically nontrivial if $Lx\neq 0$, so the circuit-level distance is
\begin{equation}
  d_{\mathrm{circ}}=\min\{\,|x| : Hx=0,\ Lx\neq 0\,\}.
  \label{eq:dcirc}
\end{equation}

\section{Proximity: A Surrogate for Circuit-Level Distance}
\label{sec:motivations}

The fault tolerance of an SEC is ultimately measured by its logical error rate (LER), but estimating the LER requires extensive sampling and decoding. The circuit-level distance is the standard analytical alternative, yet it is itself hard to compute exactly. Moreover, it is a coarse measure. Schedules with the same circuit-level distance can still differ substantially in LER, so the distance alone cannot rank them. This section introduces \emph{proximity}, a per-fault quantity on the DEM that bounds the circuit-level distance from above and can be evaluated on a small part of the DEM. \Cref{sec:motivations:proximity} defines it and relates it to the circuit-level and residual distances, and \Cref{sec:motivations:locality} explains how we choose the faults on which it is evaluated.

\subsection{Definition and Properties}
\label{sec:motivations:proximity}


Recall from \Cref{sec:background:dem} that a fault set $x\in\mathbb{F}_2^{N_f}$ is undetectable when $Hx=0$ and logically nontrivial when $Lx\neq 0$. We call a fault set with both properties an \emph{undetectable logical fault set}, and define the proximity of a single fault on this representation.

\begin{definition}[Proximity]
\label{def:dem_proximity}
The proximity of a fault $u_i\in\mathcal{U}$ is the minimum number of additional faults required to complete $u_i$ to an undetectable logical fault set,
\begin{equation}
  \prox(u_i) = \min \{\, |x|-1 \;:\; x_i=1,\ Hx=0,\ Lx\neq 0 \,\},
\end{equation}
with $\prox(u_i)=\infty$ if no such $x$ exists.
\end{definition}

Here $u_i$ is the \emph{seed}, and the other $|x|-1$ faults of a feasible $x$ form a \emph{completion} of $u_i$, which may contain faults of any type. {The term $-1$ excludes the seed itself, which every feasible $x$ contains, so $\prox(u_i)$ counts only the additional faults needed to complete $u_i$ to a logical failure, as $\sum_{j\ne i} x_j = |x|-1$.} A larger value means that the seed is farther from a logical failure, so schedules with larger proximity are preferred. Computing $\prox(u_i)$ is a minimum-weight problem over binary fault variables with parity constraints.

Two properties make proximity practically useful as a search objective. First, any set of seeds yields an \emph{upper bound} on the circuit-level distance. Specifically, for any $\mathcal{Q}\subseteq\mathcal{U}$,
\begin{equation}
  d_{\mathrm{circ}} \leq 1+\min_{u_i\in\mathcal{Q}} \prox(u_i),
  \label{eq:proximity-bound}
\end{equation}
because by \Cref{def:dem_proximity}, a seed together with a minimum-weight completion is itself an undetectable logical fault set. Equality holds if and only if $\mathcal{Q}$ contains a fault from some minimum-weight undetectable logical fault set; in particular, $\mathcal{Q}=\mathcal{U}$ gives
\begin{equation}
  d_{\mathrm{circ}} = 1+\min_{u_i\in\mathcal{U}} \prox(u_i).
  \label{eq:dem-circuit-distance}
\end{equation}

Second, this bound is \emph{tunable}. Restricting the seeds to $\mathcal{Q}$ reduces the number of completion problems, but each of them still ranges over the full DEM and is as hard as computing $d_{\mathrm{circ}}$ in the worst case. We therefore restrict the completions to $\mathcal{Q}$ as well. Let $\prox_{\mathcal{Q}}(u_i)$ denote the proximity of $u_i\in\mathcal{Q}$ computed inside $\mathcal{Q}$, i.e., \Cref{def:dem_proximity} with $x$ supported on $\mathcal{Q}$. Since a completion inside $\mathcal{Q}$ is also a completion in the full DEM, $\prox_{\mathcal{Q}}(u_i)\geq\prox(u_i)$ and
\begin{equation}
  d_{\mathrm{circ}} \;\leq\; 1+\min_{u_i\in\mathcal{Q}} \prox(u_i) \;\leq\; 1+\min_{u_i\in\mathcal{Q}} \prox_{\mathcal{Q}}(u_i).
  \label{eq:proximity-bound-restricted}
\end{equation}
Evaluating the right-hand side requires $\vert{}\mathcal{Q}\vert{}$ completion problems over $\vert{}\mathcal{Q}\vert{}$ faults each; enlarging $\mathcal{Q}$ monotonically tightens this bound, attaining $d_{\mathrm{circ}}$ if and only if $\mathcal{Q}$ contains an entire minimum-weight undetectable logical fault set.

Proximity is also closely related to the \emph{residual distance} of \citet{strikis_high-performance_2026}, which is defined on the code-capacity model and considers only hook errors, with the $X$ and $Z$ types treated separately. For the data error that a single ancilla fault leaves behind after propagating through the remaining gates of its check, the residual distance counts this fault plus the fewest data errors that complete it to a logical operator. It thus corresponds to $1+\prox(u_i)$ with the seed restricted to ancilla faults and the completion to data errors. Proximity lifts both restrictions, so its completions can also contain the measurement, idle, and correlated two-qubit faults that the decoder must handle.

\subsection{Computing Proximity on Graph-Local Fault Subsets}
\label{sec:motivations:locality}

\begin{figure}[tbp]
    \centering
    \subfloat[Local DEM graph.\label{fig:dem-incidence-graph}]{%
        \includegraphics[width=0.3\linewidth,trim={0 0 300bp 0},clip]{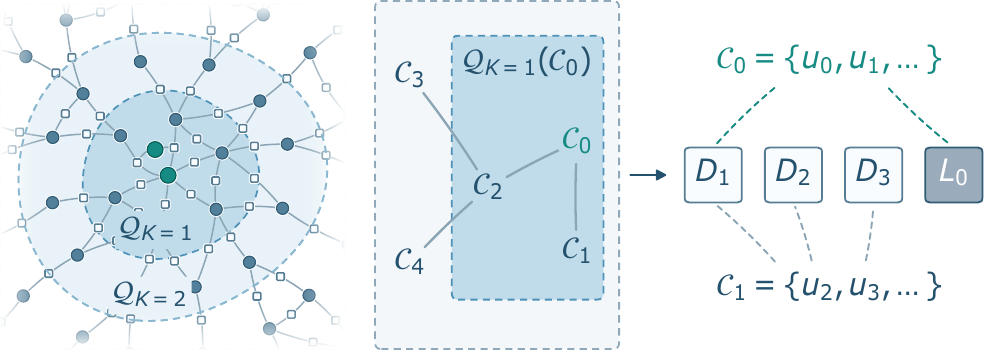}}%
    \qquad
    \subfloat[Fault grouping and expansion.\label{fig:fault-grouping}]{%
        \includegraphics[width=0.55\linewidth,trim={172bp 0 0 0},clip]{figures/proximity.pdf}}
    \caption{Proximity neighborhoods in the DEM. (a) A local region of the DEM graph, with faults as circles and detectors as squares. The highlighted faults belong to the seed class $\mathcal{C}_0$, and the shaded regions are the neighborhoods $\mathcal{Q}_K(\mathcal{C}_0)$ for $K=1$ and $K=2$. (b) Fault-class graph and neighborhood expansion. Two classes are adjacent when they contain faults that flip a common detector, e.g., $\mathcal{C}_0$ and $\mathcal{C}_1$ share $D_1$. The neighborhood of radius $K$ is the union of the classes within distance $K$ of $\mathcal{C}_0$, e.g., $\mathcal{Q}_1(\mathcal{C}_0)=\mathcal{C}_0\cup\mathcal{C}_1\cup\mathcal{C}_2$.}
    \label{fig:DEM_fault_groups}
\end{figure}

The choice of the fault subset $\mathcal{Q}$ is guided by the structure of the DEM. We call two faults \emph{adjacent} if they share a detector. A minimum-weight undetectable logical fault set is connected under this adjacency, because if it split into two parts with no detector in common, each part would be undetectable on its own, and at least one of them would flip a logical observable at strictly smaller weight. Hence a minimum-weight set of weight $w$ lies within $w-1$ adjacency steps of each of its faults, so the neighborhood of that radius around any of them contains the whole set and makes the second bound of \Cref{eq:proximity-bound-restricted} tight. We therefore take $\mathcal{Q}$ to be a small neighborhood in the DEM graph and grow it outward, as shown in \Cref{fig:dem-incidence-graph}. 

Connectivity bounds how far a neighborhood has to grow around a seed, but not which seeds it needs. Our insight for the latter comes from the translational symmetry of BB codes~\cite{bravyi_high-threshold_2024}. When the schedule respects this symmetry, the DEM inherits it, so the proximity of a fault depends only on its equivalence class under translation, and every minimum-weight undetectable logical fault set has a translate through any representative of each of its classes. The first bound, \Cref{eq:proximity-bound}, then holds with equality as soon as $\mathcal{Q}$ contains one representative of each fault class, which reduces the number of seeds from $N_f$ to the number of classes. A smaller $\mathcal{Q}$ still gives a valid upper bound, and enlarging the neighborhood tightens it by covering more classes.

The same reasoning extends, as a heuristic, to codes whose local structure is uniform but not exactly symmetric, such as quantum Tanner codes~\cite{leverrier2022quantumtannercodes,radebold2025explicit} and color codes~\cite{Bombin2007,bombin2013introduction}. Spatial boundaries and the first and last rounds break the symmetry, so a local $\mathcal{Q}$ may miss every minimum-weight fault set and the bound might not be tight; it nevertheless remains valid, and one representative neighborhood still captures the fault patterns that recur across the bulk of the circuit. \Cref{sec:methods:proximity_rank} builds the neighborhoods from fault classes of the SEC and uses the resulting scores to rank schedules; the progressive enlargement of the neighborhood is what turns proximity from a single number into a hierarchy of increasingly informative screens.



\section{The Proxima Workflow}
\label{sec:methods}

\begin{figure}[tbp]
  \centering
  \includegraphics[width=1\textwidth]{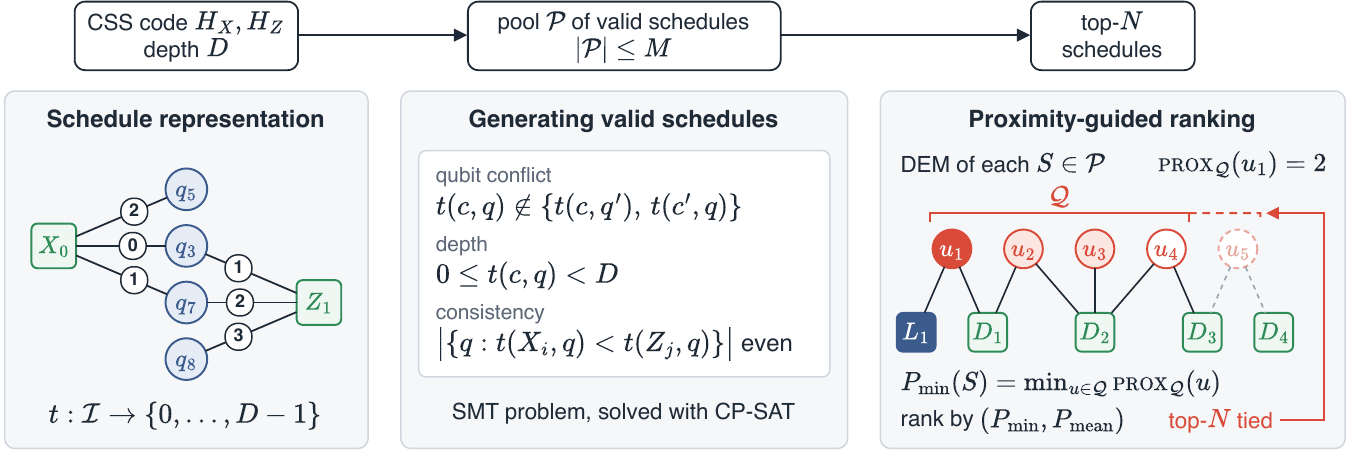}
  \caption{Overview of the Proxima workflow. The top row shows the input, the pool of valid schedules, and the returned schedules. The left panel shows how a schedule assigns a circuit layer (badge) to every ancilla--data interaction (edge), here for two overlapping checks at depth $D=4$. The middle panel lists the validity conditions, which form an SMT problem whose solutions are collected into the pool $\mathcal{P}$. The right panel shows how a schedule is scored by proximity inside a fault subset $\mathcal{Q}$ of its DEM, which is enlarged while the top-ranked schedules remain tied. In the DEM of \Cref{fig:dem-graph}, for example, the seed $u_1$ needs $u_2$ and $u_3$ to complete an undetectable logical fault set.}
  \label{fig:method-overview}
\end{figure}

Guided by the properties of proximity established in \Cref{sec:motivations}, we propose Proxima, which optimizes the SEC schedule of a CSS qLDPC code at a target circuit depth $D$. As visualized in \Cref{fig:method-overview}, Proxima describes a schedule as an assignment of circuit layers to the ancilla--data interactions of the code, synthesizes a pool of valid schedules from this description by constraint solving, and ranks the pool by proximity.
As formalized in \Cref{alg:general-schedule-evaluation}, the conditions under which a layer assignment is an executable circuit form an SMT problem, so every solution is a valid schedule by construction and the pool $\mathcal{P}$ is obtained by enumerating or sampling solutions. Each schedule in $\mathcal{P}$ is then scored on its own DEM by the proximity of the faults in a subset $\mathcal{Q}$, which, by \Cref{eq:proximity-bound-restricted}, bounds its circuit-level distance from above. Since the cost of these scores grows with $\mathcal{Q}$, Proxima starts from a small subset and enlarges it only while the top-ranked schedules remain tied, and it finally returns the top-ranked schedules.


\subsection{Schedule Representation}
\label{sec:methods:schedule-representation}
The input to schedule generation consists of the parity-check matrices $H_X$ and $H_Z$ of a CSS code and a target circuit depth $D$. For general qLDPC codes, we represent a schedule as an ordering of the nonzero entries of these matrices. Codes with additional structure, such as translational symmetry, may admit more compact representations in which equivalent stabilizers share the same local scheduling pattern. 

Specifically, each row of $H_X$ or $H_Z$ is a stabilizer check, and each nonzero entry indicates an ancilla--data interaction. We write $X_i$ and $Z_j$ for the checks given by row $i$ of $H_X$ and row $j$ of $H_Z$, $\operatorname{supp}(c)$ for the set of data qubits on which a check $c$ acts, and $(c,q)$ for the interaction between $c$ and a data qubit $q\in\operatorname{supp}(c)$, e.g., $(X_0,q_3)$ and $(Z_1,q_7)$.
Let $\mathcal{I}$ denote the set of all interactions. A candidate schedule $S$ of depth $D$ is then a layer (time slice) assignment 
\begin{equation}
  t:\mathcal{I}\to\{0,\ldots,D-1\}.
  \label{eq:layer-assignment}
\end{equation}
Listing the data qubits of a check in layer order gives its interaction sequence, e.g., $\operatorname{seq}(X_0)=(q_3,q_7,q_5,\bot)$ for $D$$\,=\,$$4$ when $t(X_0,q_3)$$\,=\,$$0$, $t(X_0,q_7)$$\,=\,$$1$, and $t(X_0,q_5)$$\,=\,$$2$, where $\bot$ marks a layer in which the ancilla idles. Relative gate ordering can therefore be read directly off the sequence, while precedence requirements become integer constraints on the layer variables, e.g., $t(X_0,q_7)<t(X_0,q_5)$.
This representation provides a unified description of candidate schedules and allows scheduling constraints to be imposed directly during the subsequent search.



For BB codes, this representation admits a more compact form that exploits their translational symmetry. By \Cref{eq:bb-defining-polynomials}, every check reaches its six data qubits through the monomials $A_1,A_2,A_3,B_1,B_2,B_3$ or their transposes, e.g., $A_1=x^3$, $A_2=y$, $A_3=y^2$, $B_1=y^3$, $B_2=x$, and $B_3=x^2$ for the $[[72,12,6]]$ code of \Cref{fig:bb-code-example}. Following \citet{bravyi_high-threshold_2024}, we let all checks of the same type apply these monomials in the same order, so that a schedule is described by one such order per check type together with the positions of its $D-6$ idle layers, as in \Cref{fig:bravyi_ancilla_view_schedule}. We call this the \emph{monomial representation} and its schedules \emph{uniform}. Since a uniform schedule consists of two such sequences regardless of the code size, the valid ones can be enumerated exhaustively. Whether non-uniform schedules can perform better remains an open question.

\subsection{Generating Valid Schedules}
\label{sec:methods:schedule-validation}

A layer assignment corresponds to an executable circuit only if it satisfies three conditions.
\begin{enumerate}
  \item \emph{Qubit conflict}: Since each interaction is a two-qubit gate between an ancilla and a data qubit, no qubit may participate in more than one interaction within the same layer. This applies both to the interactions of a single check ancilla and to those on a data qubit, which is typically shared by several $X$- and $Z$-type checks. 
  \item \emph{Depth constraint}: Every interaction must be placed within the $D$ available layers. Since a check of weight $w$, or a data qubit acted on by $w$ checks, requires $w$ distinct layers, $D$ must be at least the largest such degree in $H_X$ and $H_Z$. Smaller target depths are infeasible and rejected before the search. 
  \item \emph{Consistency constraint}: For every $X$-type and every $Z$-type check that act on common data qubits, the CNOTs acting on that shared support must be ordered with even inversion parity. This condition cancels the induced ancilla--ancilla cross terms, leaving the measured operators unchanged~\cite{zhang_optimal_2026}. 
\end{enumerate} 


Using the layer assignment $t$, these constraints can be encoded directly into an SMT problem over the integer variables $t(c,q)$. The depth constraint is the range $0\le t(c,q)<D$ of every variable. Conflict-freedom requires $t(c,q)\neq t(c,q')$ for two interactions of the same check, e.g., $t(X_0,q_3)\neq t(X_0,q_7)$, and $t(c,q)\neq t(c',q)$ for two interactions on the same data qubit, e.g., $t(X_0,q_3)\neq t(Z_1,q_3)$. For the consistency constraint, every $X$-type check $X_i$ and $Z$-type check $Z_j$ that act on common data qubits must satisfy
\begin{equation}
  \bigl|\{\,q\in\operatorname{supp}(X_i)\cap\operatorname{supp}(Z_j) : t(X_i,q)<t(Z_j,q)\,\}\bigr|\equiv 0 \pmod 2,
  \label{eq:consistency-parity}
\end{equation}
i.e., $X_i$ acts first on an even number of the shared qubits. Every solution is a valid schedule by construction, and \Cref{alg:general-schedule-evaluation} builds the pool $\mathcal{P}$ by enumerating the solutions when this is feasible and by independent randomized solves otherwise.

\subsection{Proximity-Guided Ranking}
\label{sec:methods:proximity_rank}

The schedules of the pool $\mathcal{P}$ are ranked by proximity, evaluated on the graph-local neighborhoods motivated in \Cref{sec:motivations:locality}. For each schedule, we group the faults of its DEM into classes according to their locations and roles in the SEC, and build the fault-class graph $G_{\mathrm{F}}=(V_{\mathrm{F}},E_{\mathrm{F}})$ of \Cref{fig:fault-grouping}, in which two classes are adjacent when they contain faults that share a detector.
Starting from a seed class $\mathcal{C}_0\in V_{\mathrm{F}}$, the neighborhood of radius $K$ collects the faults of all classes within graph distance $K$ of $\mathcal{C}_0$,
\begin{equation}
    \mathcal{Q}_K(\mathcal{C}_0)
    :=
    \bigcup\,\bigl\{\,\mathcal{C}\in V_{\mathrm{F}}
    :
    \operatorname{dist}_{G_{\mathrm{F}}}(\mathcal{C},\mathcal{C}_0)\leq K\,\bigr\},
    \label{eq:general-fault-neighborhoods}
\end{equation}
so that $\mathcal{Q}_0(\mathcal{C}_0)\subseteq\mathcal{Q}_1(\mathcal{C}_0)\subseteq\cdots\subseteq\mathcal{U}$, as shown in \Cref{fig:dem-incidence-graph}.

Inside a neighborhood, we use the restricted proximity $\prox_{\mathcal{Q}}$ of \Cref{sec:motivations:proximity} and abbreviate $\prox_K:=\prox_{\mathcal{Q}_K(\mathcal{C}_0)}$. The values $\prox_K(u_i)$ of the faults $u_i\in\mathcal{Q}_K(\mathcal{C}_0)$ form the \emph{proximity profile} of a schedule $S$ at level $K$. We score $S$ by the minimum of this profile,
\begin{equation}
    P_{\min}^{(K)}(S)
    :=
    \min_{u_i\in\mathcal{Q}_K(\mathcal{C}_0)}
    \prox_K(u_i),
    \label{eq:general-dem-level-score}
\end{equation}
and by the mean $P_{\mathrm{mean}}^{(K)}(S)$ of its finite values. By \Cref{eq:proximity-bound-restricted}, $1+P_{\min}^{(K)}(S)\geq d_{\mathrm{circ}}$. This bound is non-increasing in $K$ and is tight once $\mathcal{Q}_K(\mathcal{C}_0)$ contains a minimum-weight undetectable logical fault set.

Schedules are ranked by decreasing $P_{\min}^{(K)}$, with ties broken by $P_{\mathrm{mean}}^{(K)}$, and only the top-ranked ones are retained. Since the cost of each score grows with the neighborhood, we start from a small neighborhood and enlarge it only when the retained schedules remain tied, i.e., agree in both scores. In \Cref{alg:general-schedule-evaluation}, \texttt{ProximityProfile} computes the profile of a schedule and returns its two scores, and the enlargement is the outer loop. The algorithm controls $\mathcal{Q}$ by its size $s$ instead of the radius $K$, seeding it with a low-weight undetectable logical fault set, the structure analyzed in \Cref{sec:motivations:locality}, and adding faults until $|\mathcal{Q}|=s$.

For BB codes, the monomial representation makes this ranking considerably cheaper. A uniform schedule respects the translational symmetry of the code, so the proximity of a fault depends only on its class under translation, as explained in \Cref{sec:motivations:locality}. Since the number of such classes and the number of uniform schedules both remain manageable, we need not restrict the seeds to a neighborhood, and instead evaluate one representative of every class of faults that cause hook errors or idle errors.

\SetKwInput{KwParam}{Parameters}
\SetKw{KwBreak}{break}
\begin{algorithm}[tbp]
  \caption{Schedule generation and proximity-guided ranking in Proxima}
  \label{alg:general-schedule-evaluation}
  \KwIn{CSS code $(H_X,H_Z)$ of distance $d$, depth $D$ (by default the smallest feasible depth)}
  \KwParam{pool size $M$, number $N$ of returned schedules, initial subset size $s_0$,
           growth factor $\gamma>1$, maximum number $R$ of enlargements}
  \KwOut{the $N$ top-ranked schedules $\mathcal{T}$ of the pool $\mathcal{P}$}
  \BlankLine
  \tcp{Generate the pool of valid schedules as in \Cref{sec:methods:schedule-validation}}
  $\Phi\gets$ SMT problem over $t:\mathcal{I}\to\{0,\ldots,D-1\}$ with the qubit-conflict and consistency constraints\;
  \eIf{the solutions of $\Phi$ can be enumerated within the time budget}{
    $\mathcal{P}\gets$ all solutions of $\Phi$, or $M$ of them drawn uniformly at random if there are more\;
  }{
    $\mathcal{P}\gets$ the solutions returned by $M$ independent randomized solves of $\Phi$\;
  }
  \BlankLine
  \tcp{Rank the pool by proximity as in \Cref{sec:methods:proximity_rank}}
  \For{$r=0,1,\ldots,R$}{
    \lForEach{$S\in\mathcal{P}$}{$\bigl(P_{\min}(S),P_{\mathrm{mean}}(S)\bigr)\gets$ \FProx{$S,s_r$}}
    $\mathcal{T}\gets$ the $N$ schedules of $\mathcal{P}$ with the largest $P_{\min}$, ties broken by the larger $P_{\mathrm{mean}}$\;
    \lIf{the schedules of $\mathcal{T}$ are not tied}{\KwBreak}
    $s_{r+1}\gets\lceil\gamma s_r\rceil$\tcp*{enlarge the fault subset}
  }
  \KwRet{$\mathcal{T}$}\;
  \BlankLine
  \Fn{\FProx{$S,s$}}{
    $(H_S,L_S)\gets$ DEM of the $d$-round memory circuit of $S$\;
    $\mathcal{Q}\gets$ the faults of a low-weight undetectable logical fault set of $(H_S,L_S)$\;
    add faults of further such sets, then faults that flip an observable, until $|\mathcal{Q}|=s$\;
    \lForEach{$u_i\in\mathcal{Q}$}{$\prox_{\mathcal{Q}}(u_i)\gets\min\{\,|x|-1 : x_i=1,\ \operatorname{supp}(x)\subseteq\mathcal{Q},\ H_Sx=0,\ L_Sx\neq 0\,\}$}
    \KwRet{$\bigl(\min_{u_i\in\mathcal{Q}}\prox_{\mathcal{Q}}(u_i),\ \operatorname{mean}_{u_i\in\mathcal{Q}}\prox_{\mathcal{Q}}(u_i)\bigr)$}\tcp*{mean over the finite values}
  }
\end{algorithm}

\subsection{Implementation and Complexity}
\label{sec:method:implementation}

We implement Proxima in Python, following \Cref{alg:general-schedule-evaluation}. Stim~\cite{gidney2021stim} compiles each schedule into a $d$-round memory circuit and extracts its DEM. Since the SMT problem of the schedule pool and the completion problems of the proximity profiles all have finite domains, we solve them with the CP-SAT solver of Google OR-Tools~\cite{ortools}. The pool keeps one schedule among solutions that order the CNOTs identically on every qubit. The low-weight fault set that starts $\mathcal{Q}$ is found by exhaustive search up to weight~3 and by the search routine of Stim otherwise. Each completion problem runs under a time limit $t_q$, and a profile is ranked only if all of its problems finish, so every ranked value is a proven optimum. Profiles of different schedules are computed in parallel.

The schedule pool is generated once under a fixed time budget. Without the consistency constraint, finding a valid schedule is an edge coloring of the bipartite Tanner graph~\cite{durazo_rocha_local_2026}, which takes polynomial time~\cite{cole2001edge}, but with the consistency constraint included, no polynomial-time algorithm is currently known. The hardness of Proxima lies in the proximity profiles since, by \Cref{eq:dem-circuit-distance}, computing proximity over the full DEM is as hard as computing $d_{\mathrm{circ}}$, which generalizes the NP-hard minimum distance of a binary linear code~\cite{vardy1997intractability}. Restricting each completion problem to $\mathcal{Q}$ bounds its cost by $2^{O(s)}$ instead of $2^{O(N_f)}$, and by $O(s^{w})$ if its optimum has weight $w$, at the price of an upper bound on the proximity, as in \Cref{eq:proximity-bound-restricted}. Proxima solves $|\mathcal{P}|\sum_r s_r$ such problems, where $s_r$ is the subset size in pass $r$ of the outer loop. Since $s_r$ grows by a factor of at least $\gamma>1$ per pass, $\sum_r s_r\leq\tfrac{\gamma}{\gamma-1}s_{\max}$ for the largest size $s_{\max}$ reached, so the total solver time is at most $\tfrac{\gamma}{\gamma-1}|\mathcal{P}|\,s_{\max}\,t_q$, which is linear in the pool size.

\section{Evaluation}\label{sec:evaluation}


We evaluate Proxima against to address the following research questions.

\textbf{RQ1.} \emph{Does Proxima find SEC schedules with a lower LER than existing automated schedulers across qLDPC code families?}

\textbf{RQ2.} \emph{On BB codes, do the uniform schedules found in the monomial representation outperform IBM's reference schedule and the automated baselines?}

\textbf{RQ3.} \emph{How much does each component of the proximity-guided ranking contribute to the final fault-tolerant performance?}

\textbf{RQ4.} \emph{What is the computational cost of Proxima, and how does it grow with the code size?}

\textbf{RQ5.} \emph{Do the conclusions hold under other noise models and decoders?}

\subsection{Experimental Settings}
\label{sec:eval:settings}

\paragraph{Benchmarks}
We use the eight CSS codes listed in \Cref{tab:benchmarks}, which span six code families, check weights from 4 to 8, and sizes from 24 to 625 data qubits. All of them are selected from prior related work for fair comparison~\cite{bravyi_high-threshold_2024,strikis_high-performance_2026,zhang_optimal_2026,liu_alphasyndrome_2026,viszlai_prophunt_2026}.

\begin{table}[t]
  \centering
  \caption{Benchmark qLDPC codes.}
  \label{tab:benchmarks}
\footnotesize
\setlength{\tabcolsep}{4pt}
\renewcommand{\arraystretch}{1.2}
\begin{tabular}{|l|l|c|c|}
  \hline
  \textbf{Code} & \textbf{Code family} & $[[n,k,d]]$ & \textbf{Stabilizer weight} \\
  \hline
  \texttt{HexCC61}   & Hexagonal color code  & $[[61,1,9]]$    & 4, 6    \\ \hline
  \texttt{HyperCC24} & Hyperbolic color code & $[[24,8,4]]$    & 6, 8    \\ \hline
  \texttt{RQT108}    & Random quantum Tanner code    & $[[108,18,4]]$  & 6       \\ \hline
  \texttt{LP39}      & Lifted product code   & $[[39,3,3]]$    & 4, 5, 6 \\ \hline
  \texttt{HGP52}     & \multirow{2}{*}{Hypergraph product code} & $[[52,4,4]]$ & 4, 5, 6 \\ \cline{1-1}\cline{3-4}
  \texttt{HGP625}    &                       & $[[625,25,8]]$  & 7       \\ \hline
  \texttt{BB108}     & \multirow{2}{*}{Bivariate bicycle code}  & $[[108,8,10]]$ & 6 \\ \cline{1-1}\cline{3-4}
  \texttt{BB144}     &                       & $[[144,12,12]]$ & 6       \\ \hline
\end{tabular}

\end{table}

\paragraph{Baselines}
We compare Proxima with the automated schedulers of \Cref{tab:taxonomy}, i.e., ASC~\cite{zhang_optimal_2026}, LRC~\cite{strikis_high-performance_2026}, AlphaSyndrome~\cite{liu_alphasyndrome_2026}, and PropHunt~\cite{viszlai_prophunt_2026}, and on BB codes also with IBM's depth-7 reference schedule~\cite{bravyi_high-threshold_2024}. Each baseline runs its public implementation with default settings under a time limit of 10 hours per code, and every method is represented by its best-performing schedule. Proxima uses a pool of $M=1000$ schedules ($M=50$ on \texttt{HGP625}), an initial subset size $s_0=8$, and returns $N=3$ schedules.
All experiments run on a dual-socket server equipped with two 96-core AMD EPYC 9684X processors and 1.5 TiB of RAM.

\paragraph{Noise model and decoder}
\label{sec:method:ler_simu}
All schedules are compiled into memory circuits and sampled with Stim~\cite{gidney2021stim} under the standard circuit-level depolarizing noise model with a single physical error rate $p$~\cite{bravyi_high-threshold_2024}. Each CNOT, ancilla preparation, and ancilla measurement fails with probability $p$, and so does each data or ancilla qubit that idles during a CNOT layer and each data qubit that idles during ancilla measurement and reset. By default, we set the physical error rate to $p=10^{-3}$. Syndromes are decoded with the BP-OSD decoder~\cite{Panteleev2021degeneratequantum,roffe_decoding_2020,Roffe_LDPC_Python_tools_2022}, \unsure{using min-sum BP with up to $10^4$ iterations followed by order-7 combination-sweep OSD}. We also investigate the evaluated methods' performance with varying error rates, noise models, and decoders.



\paragraph{Metrics}
We mainly report the circuit depth $D$, circuit-level distance $d_{\mathrm{circ}}$, and LER as the key metrics. $D$ refers to the number of CNOT layers per syndrome extraction cycle. $d_{\mathrm{circ}}$ is reported as an upper bound, namely the weight of the lightest undetectable logical fault set found by \texttt{tropidist}~\cite{cheng_certifying_2026}, which evaluates the distance on the DEM with rank-based dynamic programming and weight-bounded search. LER is defined as the per-cycle logical error rate $p_L=1-(1-P_L)^{1/N_c}$, where $P_L$ is the fraction of shots with a logical failure in a memory experiment of $N_c=d$ noisy cycles. Each data point uses up to $10^6$ shots, and we report 95\% Wilson confidence intervals transformed to the per-cycle LER.

\subsection{RQ1: Comparison with Existing Schedulers}
\label{sec:eval:rq1}

\begin{table}[t]
  \centering
      \caption{Benchmarking of Proxima against other SOTA scheduling methods. The table summarizes the best-performing schedule for each method, along with the corresponding circuit depth $D$, an upper bound on the circuit-level distance $d_{\mathrm{circ}}$, and the per-cycle logical error rate (LER) at a representative physical error rate $p=10^{-3}$. Shaded cells mark the best value for each code, whose parameters are listed in \Cref{tab:benchmarks}.}
      \label{tab:result_summary}
      \centering
%
\newcommand{\lerval}[2]{$#1{\times}10^{#2}$}
\newcommand{\dub}[1]{$\leq #1$}
\newcommand{\best}[2][codegreen]{\cellcolor{#1}{#2}}
\definecolor{codegreen}{HTML}{EBF2EF}
\definecolor{codeblue}{HTML}{E8EFF8}
\footnotesize
\setlength{\tabcolsep}{2.4pt}
\setlength{\doublerulesep}{1pt}
\renewcommand{\arraystretch}{1.3}
\begin{tabular}{|l||c|c|c||c|c|c||c|c|c||c|c|c||c|c|c|}
  \hline
  \multirow{2}{*}{\textbf{Code}}
    & \multicolumn{3}{c||}{\textbf{ASC}}
    & \multicolumn{3}{c||}{\textbf{LRC}}
    & \multicolumn{3}{c||}{\textbf{AlphaSyndrome}}
    & \multicolumn{3}{c||}{\textbf{PropHunt}}
    & \multicolumn{3}{c|}{\textbf{Proxima}} \\
  \cline{2-16}
  & $D$ & $d_{\mathrm{circ}}$ & LER & $D^{*}$ & $d_{\mathrm{circ}}$ & LER &
  $D$ & $d_{\mathrm{circ}}$ & LER & $D$ & $d_{\mathrm{circ}}$ & LER &
  $D$ & $d_{\mathrm{circ}}$ & LER \\
  \hline
  \texttt{\scriptsize{HexCC61}}
  & \best{6} & \best{\dub{6}} & \lerval{5.6}{-5}  
  & 10 & \dub{5} & \lerval{1.1}{-4}  
  & 20 & \dub{5} & \lerval{8.7}{-4}  
  & 14 & \dub{5} & \lerval{3.5}{-4}  
  & \best{6} & \best{\dub{6}} & \best{\lerval{2.2}{-5}}  
  \\
  \hline
  \texttt{\scriptsize{HyperCC24}}
  & \best{8} & \dub{2} & \lerval{1.8}{-2}  
  & 10 & \dub{2} & \lerval{2.2}{-2}
  & 18 & \dub{2} & \lerval{4.9}{-2}
  & 16 & \dub{2} & \lerval{4.5}{-2}
  & \best{8} & \best{\dub{3}} & \best{\lerval{1.0}{-2}}  
  \\
  \hline
  \texttt{\scriptsize{RQT108}}
  & 8 & \dub{2} & \lerval{5.5}{-3}  
  & 11 & \dub{2} & \lerval{1.0}{-2}
  & \multicolumn{3}{c||}{*timeout}  
  & 30 & \best{\dub{4}} & \lerval{8.8}{-2}
  & \best{7} & \dub{3} & \best{\lerval{4.0}{-3}}
  \\
  \hline
  \texttt{\scriptsize{LP39}}
  & \best{6} & \dub{2} & \lerval{1.3}{-3}  
  & 8 & \dub{2} & \lerval{2.4}{-3}
  & 17 & \best{\dub{3}} & \lerval{5.8}{-3}
  & 16 & \best{\dub{3}} & \lerval{6.2}{-3}
  & \best{6} & \best{\dub{3}} & \best{\lerval{6.1}{-4}}  
  \\
  \hline
  \texttt{\scriptsize{HGP52}}
  & \best{7} & \best{\dub{4}} & \lerval{2.4}{-3}  
  & 8 & \best{\dub{4}} & \lerval{3.0}{-3}
  & 18 & \best{\dub{4}} & \lerval{1.7}{-2}
  & 13 & \best{\dub{4}} & \lerval{8.0}{-3}
  & \best{7} & \best{\dub{4}} & \best{\lerval{1.2}{-3}}  
  \\
  \hline
  \texttt{\scriptsize{HGP625}}
  & 10 & \best{\dub{8}} & \lerval{9.4}{-5}  
  & 10 & \best{\dub{8}} & \lerval{1.5}{-4}
  & \multicolumn{3}{c||}{*timeout}  
  & 18 & \best{\dub{8}} & \lerval{1.5}{-3}
  & \best{8} & \best{\dub{8}} & \best{\lerval{3.8}{-5}}
  \\
  \hline
  \texttt{\scriptsize{BB108}}
  & \best[codeblue]{7} & \dub{7} & \lerval{1.0}{-6}  
  & \best[codeblue]{7} & \dub{8} & \lerval{1.0}{-6}  
  & 27 & \dub{6} & \lerval{8.7}{-4}  
  & 15 & \dub{7} & \lerval{6.0}{-5}  
  & 8 & \best[codeblue]{\dub{9}} & \best[codeblue]{\lerval{8.0}{-7}}  
  \\
  \hline
  \texttt{\scriptsize{BB144}}
  & \best[codeblue]{7} & \dub{8} & \best[codeblue]{\lerval{1.3}{-7}}  
  & \best[codeblue]{7} & \dub{8} & \lerval{4.2}{-7}  
  & 30 & \dub{7} & \lerval{7.3}{-4}  
  & 14 & \dub{7} & \lerval{1.5}{-5}  
  & 8 & \best[codeblue]{\dub{10}} & \lerval{1.7}{-7}  
  \\
  \hline
\end{tabular}

      \begin{flushleft}
        \small
        $^{*}$LRC's depths count ticks of the steady-state period rather than CNOT layers.
      \end{flushleft}
\end{table}

\begin{figure}[t]
    \centering
    \includegraphics[width=\textwidth]{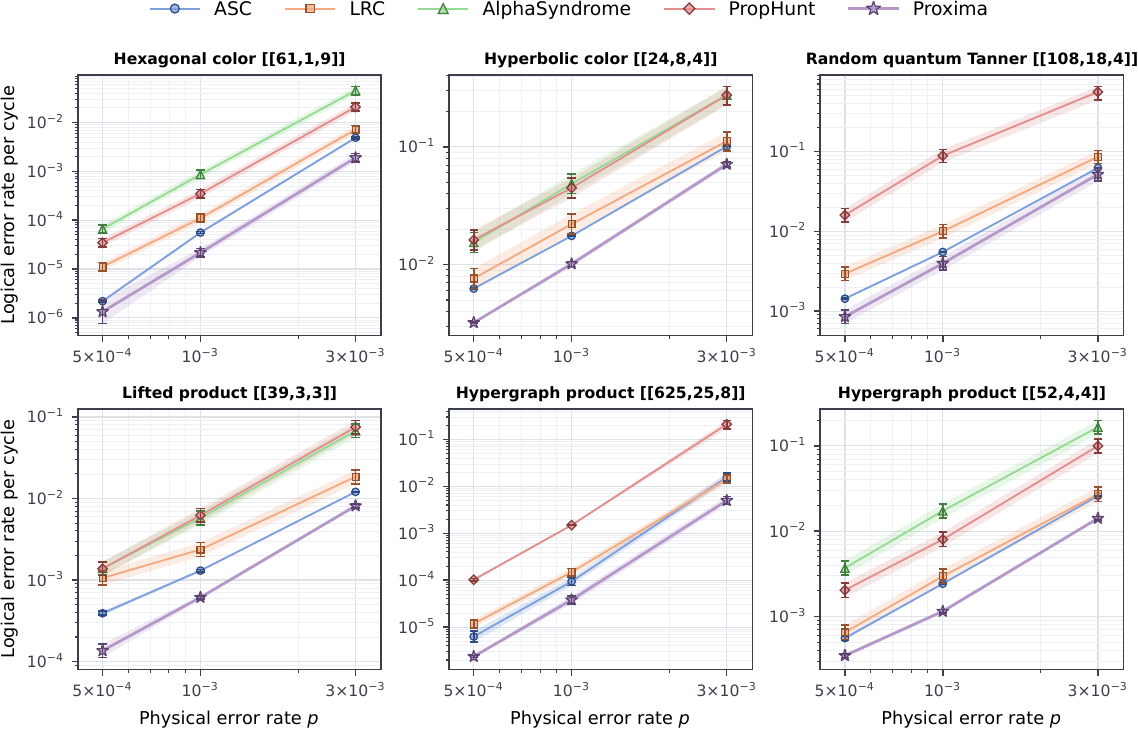}
    \caption{Per-cycle logical error rate versus physical error rate for schedules from ASC, AlphaSyndrome, PropHunt, LRC, and our proposed Proxima, under the noise model and decoder of \Cref{sec:eval:settings}. AlphaSyndrome timed out on \texttt{RQT108} and \texttt{HGP625}. Shaded bands indicate 95\% confidence intervals.}
    \label{fig:all-codes-evaluation}
\end{figure}

\Cref{tab:result_summary} reports, for every method, the depth, an upper bound on the circuit-level distance, and the LER at $p=10^{-3}$ of its schedule, and \Cref{fig:all-codes-evaluation} shows the LER of the six non-BB codes at $p\in\{5\times10^{-4},10^{-3},3\times10^{-3}\}$.
Proxima attains the lowest LER on all six codes at every physical error rate. At $p=10^{-3}$, its LER is lower than that of ASC, the strongest baseline on every code, by a factor of \unsure{1.4 to 2.6}, lower than that of LRC by a factor of 2.2 to 5.1, and lower than those of PropHunt and AlphaSyndrome by a factor of 4.4 to 40. \unsure{Among the 18 combinations of code and physical error rate, the 95\% confidence intervals of Proxima and the best baseline overlap only for \texttt{HexCC61} at $p=5\times10^{-4}$ and \texttt{RQT108} at $p=3\times10^{-3}$.}

Part of this gap comes from depth. The circuits of AlphaSyndrome and PropHunt are 1.9--4.3$\times$ deeper than those of Proxima, e.g., 30 against 7 layers on \texttt{RQT108}, and those of LRC take one to four more time steps per cycle. \blabla{Under our noise model every additional layer adds idle errors, which is why the two methods with the deepest circuits have the highest LERs even though AlphaSyndrome optimizes the LER directly.} ASC separates the effect of the gate order from that of depth. It has the same depth as Proxima on four codes, where the LER of Proxima is still lower by a factor of \unsure{1.7 to 2.6}, and on \texttt{RQT108} and \texttt{HGP625} Proxima is in addition one and two layers shallower\unsure{, since the solver of ASC does not reach the minimal depth there within its time limit}.

The circuit-level distance does not explain the remaining difference. On \texttt{HexCC61} and \texttt{HGP52}, ASC and Proxima have the same depth and the same bound on $d_{\mathrm{circ}}$, yet their LERs differ by \unsure{more than a factor of two}. On \texttt{RQT108}, PropHunt has the largest bound, 4 against 3 for Proxima, but needs 30 layers and has a 22$\times$ higher LER. \blabla{A surrogate therefore has to resolve schedules beyond their minimum-weight failure and to compare them at a fixed, low depth, which is how Proxima uses the proximity profile.}

\begin{takeaways}[title={RQ1 Summary}]
    Proxima achieves the lowest LER across all benchmarks---cutting LER by 1.4--2.6$\times$ over ASC and up to $40\times$ over deeper schedulers---even when matched in circuit depth and $d_{\mathrm{circ}}$.
\end{takeaways}

\subsection{RQ2: Uniform Schedules for BB Codes}
\label{sec:eval:bb}

\begin{figure}[tbp]
  \centering
  \includegraphics[width=.8\textwidth]{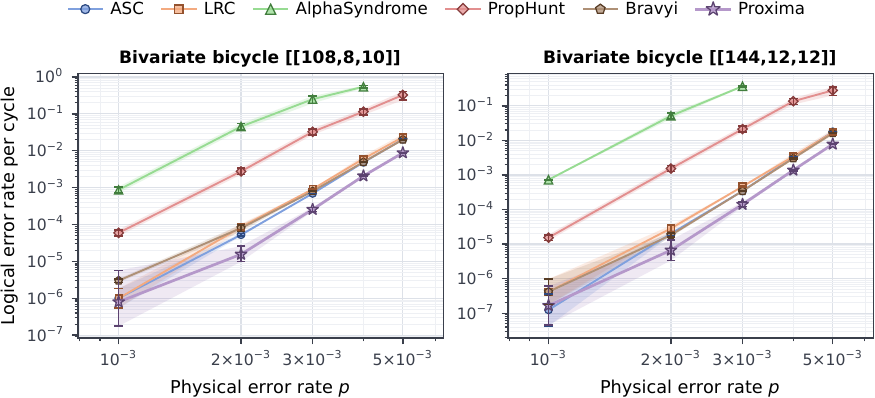}
  \caption{Per-cycle logical error rate versus physical error rate on \texttt{BB108} and \texttt{BB144} for the schedules from ASC, LRC, AlphaSyndrome, PropHunt, \citet{bravyi_high-threshold_2024}, and the depth-8 schedule of Proxima. AlphaSyndrome is not shown at $p=5\times10^{-3}$ on \texttt{BB108} and at $p\geq4\times10^{-3}$ on \texttt{BB144}, where all of its shots fail and its per-cycle LER saturates at one. Shaded bands indicate 95\% confidence intervals.}
  \label{fig:bb-ler}
\end{figure}

For BB codes, Proxima restricts the search to the uniform schedules of the monomial representation introduced in \Cref{sec:methods:schedule-representation}, \unsure{and the valid ones are enumerated exhaustively rather than sampled}. \Cref{tab:bravyi-proximity-schedules} lists the depth-7 and depth-8 schedules selected by Proxima next to the depth-7 reference schedule of \citet{bravyi_high-threshold_2024}, and \Cref{fig:bb-ler} compares the depth-8 schedule with the reference schedule and the automated baselines for $p$ from $10^{-3}$ to $5\times10^{-3}$. The same monomial sequences are used on both codes.

\begin{table}[tbp]
  \caption{Example SEC schedules generated by Proxima compared with that of \citet{bravyi_high-threshold_2024} for BB codes. A dash (--) denotes an idle layer.}
  \label{tab:bravyi-proximity-schedules}
  \definecolor{xcheck}{HTML}{C2701E}
  \definecolor{zcheck}{HTML}{2E8B57}
  \begin{center}
    \small
    \begin{tabular}{@{}lcl}
      \toprule
      \textbf{Method} & \textbf{Depth} & \textbf{Schedule in monomial representation} \\
      \midrule
      \multirow{2}{*}{\citet{bravyi_high-threshold_2024}} & \multirow{2}{*}{7}
        & \cellcolor{xcheck!15}\sector{X} \slot{\text{--}}\slot{A_2}\slot{B_2}\slot{B_1}\slot{B_3}\slot{A_1}\slot{A_3} \\
        & & \cellcolor{zcheck!15}\sector{Z} \slot{A_1^{\mathsf T}}\slot{A_3^{\mathsf T}}\slot{B_1^{\mathsf T}}\slot{B_2^{\mathsf T}}\slot{B_3^{\mathsf T}}\slot{A_2^{\mathsf T}}\slot{\text{--}} \\
      \midrule
      \multirow{2}{*}{Proxima} & \multirow{2}{*}{7}
        & \cellcolor{xcheck!15}\sector{X} \slot{A_1}\slot{A_2}\slot{B_2}\slot{B_1}\slot{B_3}\slot{\text{--}}\slot{A_3} \\
        & & \cellcolor{zcheck!15}\sector{Z} \slot{\text{--}}\slot{A_3^{\mathsf T}}\slot{B_1^{\mathsf T}}\slot{B_2^{\mathsf T}}\slot{B_3^{\mathsf T}}\slot{A_1^{\mathsf T}}\slot{A_2^{\mathsf T}} \\
      \midrule

      \multirow{2}{*}{Proxima} & \multirow{2}{*}{8}
        & \cellcolor{xcheck!15}\sector{X} \slot{A_3}\slot{B_1}\slot{B_2}\slot{\text{--}}\slot{A_1}\slot{\text{--}}\slot{B_3}\slot{A_2} \\
        & & \cellcolor{zcheck!15}\sector{Z} \slot{A_2^{\mathsf T}}\slot{B_3^{\mathsf T}}\slot{\text{--}}\slot{A_1^{\mathsf T}}\slot{\text{--}}\slot{B_1^{\mathsf T}}\slot{B_2^{\mathsf T}}\slot{A_3^{\mathsf T}} \\
      \bottomrule
    \end{tabular}
  \end{center}
\end{table}

Compared with the reference schedule, the depth-8 schedule of Proxima lowers the LER by a factor of 2.3 to 5.1 on \texttt{BB108} and of 2.1 to 2.7 on \texttt{BB144}, i.e., by 52\% to 81\%. Both factors grow as $p$ decreases from $5\times10^{-3}$ to $2\times10^{-3}$, which is consistent with the selected schedule suppressing low-weight failure mechanisms rather than rescaling the error rate uniformly. At $p=10^{-3}$, where few failures are observed, the factors are 3.8 and 2.5 and the confidence intervals overlap.

Against the automated baselines, which do not restrict their search to uniform schedules, the depth-8 schedule also has the lowest LER for $p\geq2\times10^{-3}$. It is lower than those of ASC and LRC by a factor of 2.3 to 5.5 on \texttt{BB108} and of 2.2 to 4.3 on \texttt{BB144}, and lower than those of PropHunt and AlphaSyndrome, whose circuits have 14 to 30 layers, by a factor of more than 35. At $p=10^{-3}$, the estimates of ASC, LRC, and Proxima rest on a few failures and their intervals overlap, with ASC nominally the lowest on \texttt{BB144}.

The depth-8 schedule wins although it has one more layer than the schedules of ASC, LRC, and \citet{bravyi_high-threshold_2024}, so its gate order outweighs the additional idle errors. Proxima's depth-7 schedule is nominally worse as well, at $2.5\times10^{-7}$ against $1.7\times10^{-7}$ on \texttt{BB144} for $p=10^{-3}$. The distance bound of the depth-8 schedule in \Cref{tab:result_summary} is also the highest on both codes, $\leq 9$ on \texttt{BB108} and $\leq 10$ on \texttt{BB144}, whereas no automated baseline exceeds 8. \unsure{For the reference schedule, \mbox{\citet{bravyi_high-threshold_2024}} report bounds of 8 and 10.}
\Cref{sec:eval:sensitivity} repeats the comparison on \texttt{BB108} under other noise models and a second decoder.

\begin{takeaways}[title={RQ2 Summary}]
    Exploiting translational symmetry via monomials, Proxima discovers uniform depth-8 BB schedules that cut LER by 2.1--5.1$\times$ over IBM's depth-7 reference and by over $2.2\times$ against all automated baselines.
\end{takeaways}

\subsection{RQ3: Ablation Study}
\label{sec:eval:ablation}

We study how circuit depth $D$, fault-subset size $s$, and profile aggregation affect schedule selection on \texttt{LP39} and \texttt{HyperCC24}. For each code, we randomly generate 100 schedules at each of two consecutive depths and evaluate their LERs under the settings of \Cref{sec:eval:settings}. We progressively filter the lower-depth pool as the fault-subset size $s$ increases. At each stage, we rank the remaining schedules by proximity. If multiple schedules tie for the best rank, we retain this top proximity class; otherwise, we retain the top half of the ranked schedules, rounding up when necessary. On \texttt{LP39}, the retained sets contain 50, 25, and 13 candidates at $s=8$, 16, and 32, respectively. On \texttt{HyperCC24}, they contain 30 and 10 candidates at $s=8$ and 16. The retained sets are nested as $s$ increases. We compare the three schedules selected by $P_{\mathrm{mean}}$ alone, $P_{\min}$ alone, and Proxima from the final retained set.

\begin{figure}[tbp]
  \centering
  \includegraphics[width=0.9\linewidth]{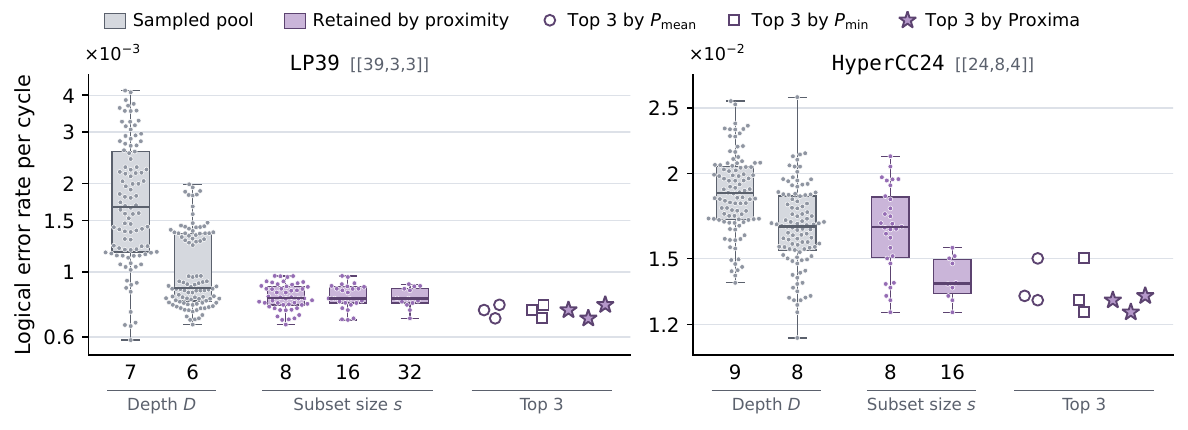}
  \caption{Ablation study for how different proximity-guided granularities in scheduling optimization contribute to the final gains. Each dot is one schedule. Boxes span the interquartile range with a line at the median, and whiskers reach the minimum and maximum. Grey columns show the 100 schedules sampled at each depth $D$, and purple columns the schedules of the lower-depth pool that proximity ranking retains at fault-subset size $s$. The markers are the three schedules selected by $P_{\mathrm{mean}}$ alone, by $P_{\min}$ alone, and by Proxima.}
  \label{fig:ablation_study}
\end{figure}

As shown in \Cref{fig:ablation_study}, reducing the circuit depth by one decreases the maximum LER by 52.10\% on \texttt{LP39}, but slightly increases it by 1.35\% on \texttt{HyperCC24}. Proximity-based ranking further narrows the LER range of the retained candidates. On \texttt{LP39}, a small fault-subset size of $s=8$ already decreases the maximum LER of the lower-depth pool by 51.23\%, comparable to depth reduction. Increasing $s$ to 16 leaves it unchanged, and $s=32$ lowers it by a further 6.27\%. On \texttt{HyperCC24}, $s=8$ decreases the maximum LER by 18.18\% but leaves the median essentially unchanged, whereas $s=16$ decreases the maximum by a further 26.71\% and the median by 17.37\%. The fault-subset size needed to separate the candidates thus depends on the code. Ranking the retained candidates by the mean or the minimum of the proximity profile alone selects the same three schedules as Proxima on \texttt{LP39}, but on \texttt{HyperCC24} each admits a schedule with a 13.57\% higher LER than any selected by Proxima. Combining the two statistics, as in Proxima, selects the three lowest-LER retained candidates on both codes.

\begin{takeaways}[title={RQ3 Summary}]
    Progressive neighborhood enlargement resolves degeneracies on denser codes, while combining worst-case ($P_{\min}$) and average ($P_{\mathrm{mean}}$) proximity consistently isolates the lowest-LER schedules from the candidate pool.
\end{takeaways}

\subsection{RQ4: Scalability Analysis}
\label{sec:eval:scalability}

A Proxima run spends its time on generating the schedule pool and on ranking it by proximity. Both parts only solve constraint problems on the code and its DEM, and nothing is sampled or decoded before the retained schedules are simulated. In contrast, as summarized in \Cref{tab:taxonomy}, AlphaSyndrome samples and decodes at every rollout, and PropHunt solves MaxSAT instances on sampled DEM subgraphs in every iteration. AlphaSyndrome already fails to finish on \texttt{RQT108} and \texttt{HGP625}, the two largest non-BB codes.


\begin{figure}[tbp]
    \centering
    \includegraphics[width=0.6\linewidth]{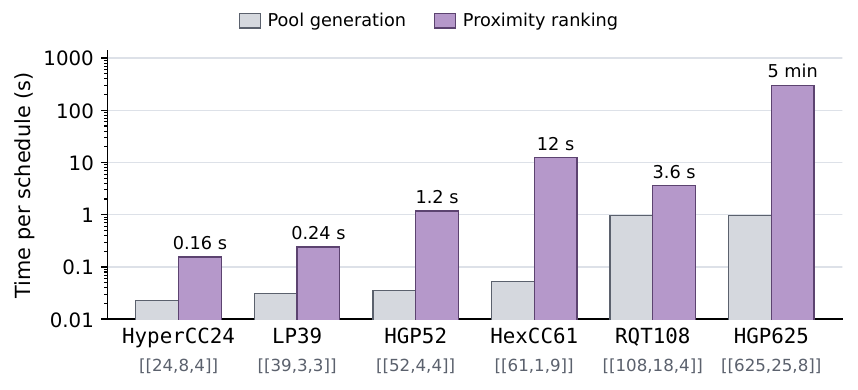}
    \caption{Wall-clock time per schedule of the two parts of a Proxima run with eight workers, on the six non-BB codes ordered by their number of data qubits. Each bar is the time of one part divided by the pool size $M$, which is 50 for \texttt{HGP625} and 1000 otherwise.}
    \label{fig:scalability_analysis}
\end{figure}

We time Proxima on the six non-BB codes with $M=1000$ schedules and $s=16$ faults without enlargement. \texttt{HexCC61} and \texttt{HGP625} use $s=8$, and \texttt{HGP625} only $M=50$. \Cref{fig:scalability_analysis} reports the wall-clock time per schedule. A run with $M=1000$ takes between 3~minutes on \texttt{HyperCC24} and 3.4~hours on \texttt{HexCC61}, and the 50 schedules of \texttt{HGP625} take 4.2~hours. Proximity ranking accounts for 79\% to 99.7\% of this time, whereas generating the pool costs at most one second per schedule. The time to rank one schedule does not follow the number of qubits alone. It grows from 0.16\,s on \texttt{HyperCC24} to 3.6\,s on \texttt{RQT108}, yet \texttt{HexCC61} needs 12\,s and \texttt{HGP625} 5~minutes. The completion problems are not the cause, since each involves only the $s$ faults of the subset. The cost lies in the DEM, which grows with both $n$ and $d$, and in the search over it for the fault set that starts the subset, described in \Cref{sec:method:implementation}. This search dominates on these two codes and times out for 21 of the 1000 schedules of \texttt{HexCC61}, which are left unranked. Since schedules are ranked independently, a run scales linearly with $M$, which is why \texttt{HGP625} is limited to 50 schedules.

\begin{takeaways}[title={RQ4 Summary}]

    Operating purely via constraint solving without in-loop simulation, Proxima screens 1,000 schedules in minutes to hours with linear pool scaling, reaching 625-qubit codes where simulation-driven schedulers time out.
\end{takeaways}

\subsection{RQ5: Sensitivity Analysis}
\label{sec:eval:sensitivity}

\begin{figure}[tbp]
  \centering
  \subfloat[Varying noise models.\label{fig:sensitivity-noise}]{%
    \includegraphics[height=0.288\linewidth]{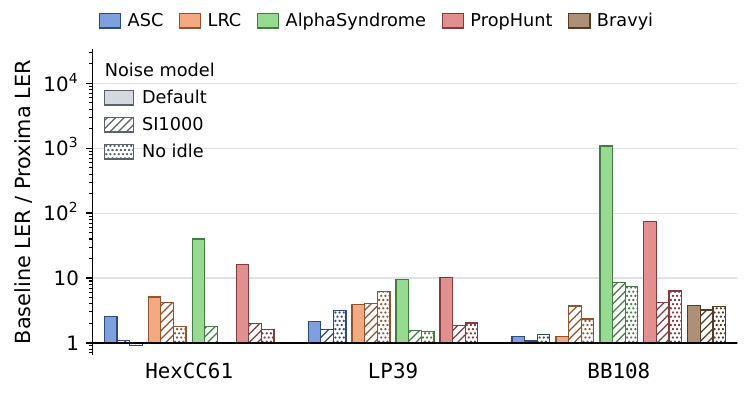}}%
  \hfill
  \subfloat[Varying decoders.\label{fig:sensitivity-decoder}]{%
    \includegraphics[height=0.288\linewidth]{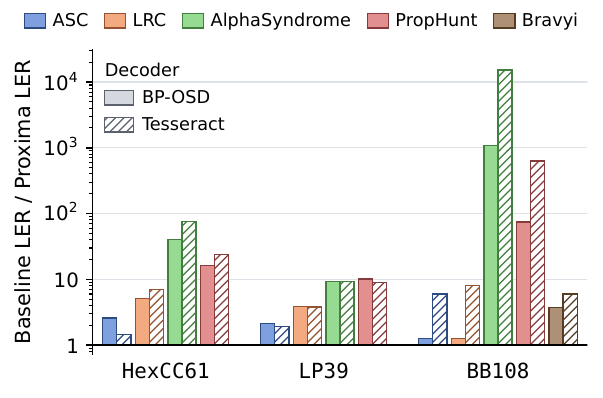}}
  \caption{Sensitivity of relative LER performance to (a) the noise model and (b) the decoder at $p=10^{-3}$. Each bar shows the LER of a baseline schedule normalized to that of Proxima for the same code and configuration; values greater than $1$ indicate lower LER for Proxima.}
  \label{fig:sensitivity}
\end{figure}

To evaluate whether our comparative findings generalize across noise assumptions and decoders, we take the fixed schedules produced by each method for \texttt{HexCC61}, \texttt{LP39}, and \texttt{BB108} and re-evaluate them under alternative configurations without rerunning schedule optimization. All other experimental parameters follow \Cref{sec:eval:settings}. \Cref{fig:sensitivity} reports the LER of each baseline normalized to that of Proxima at the representative physical error rate of \unsure{$p=10^{-3}$}.

\subsubsection{Sensitivity to Noise Models}
\label{sec:eval:sens-noise}

Beyond our default noise model, we evaluate two alternative noise regimes: the superconducting-inspired SI1000 model~\cite{gidney2021honeycomb}, in which idling error rates are ten times lower than gate error rates, and an idle-free variant of the default model that omits idling errors altogether (matching the evaluation setup of PropHunt~\cite{viszlai_prophunt_2026}). Both settings attenuate the idling noise that heavily penalizes the deeper schedules generated by AlphaSyndrome and PropHunt under the default model.

Under SI1000, Proxima maintains the lowest LER across all three codes, though the performance margins shift noticeably. The advantage over deeper schedules narrows sharply---for example, dropping from $40\times$ to $1.8\times$ against AlphaSyndrome and from $16\times$ to $2.0\times$ against PropHunt on \texttt{HexCC61}---confirming that their degradation in RQ1 stems primarily from idling noise. Meanwhile, Proxima retains an approximately $4\times$ lower LER than LRC across all three codes. Against ASC, the margin narrows from \unsure{$2.6\times$} to $1.1\times$ on \texttt{HexCC61} and from \unsure{$2.1\times$} to $1.6\times$ on \texttt{LP39}, with overlapping confidence intervals between the two methods on \texttt{HexCC61} and \texttt{BB108}. \blabla{Because proximity counts fault mechanisms uniformly rather than weighting them by error probability, it does not explicitly exploit the heterogeneous error rates of SI1000.}

When idling errors are removed entirely, circuit depth no longer incurs idling penalties, and Proxima's advantage over the deeper schedules narrows to 1.0--7.3$\times$. On \texttt{HexCC61}, Proxima, ASC, and AlphaSyndrome achieve LERs within roughly $10\%$ of one another with overlapping confidence intervals, with ASC attaining a nominally lower point estimate. On \texttt{LP39}, Proxima continues to outperform every baseline by 1.5--6.3$\times$, and its advantage over ASC even widens from \unsure{$2.1\times$} to $3.1\times$. On \texttt{BB108}, Proxima again achieves the lowest point LER, though limited logical failure counts result in confidence intervals that overlap with those of ASC, LRC, and the reference schedule.

\subsubsection{Sensitivity to Decoders}
\label{sec:eval:sens-decoder}

Under the default noise model at $p=10^{-3}$, we also evaluate all schedules using Tesseract~\cite{beni2025tesseractdecoder}, a near-optimal most-likely-error decoder. We run it on $10^7$ shots per data point with beam width 15, a queue limit of 200{,}000, beam climbing, no detector revisiting, and error merging, and its low-confidence predictions count as failures only when incorrect. Whereas AlphaSyndrome explicitly optimizes schedules with BP-OSD inside its search loop~\cite{liu_alphasyndrome_2026}, the other approaches operate in a decoder-agnostic manner; this experiment therefore tests whether our comparative gains transfer across decoders. On \texttt{BB108}, it further tests whether Proxima's depth-8 schedule preserves its advantage over depth-7 counterparts under near-optimal decoding.

While Tesseract reduces the absolute LER of every schedule, Proxima retains the lowest LER across all three codes. The relative ordering of baselines on \texttt{HexCC61} and \texttt{LP39} mirrors that under BP-OSD, except that PropHunt and AlphaSyndrome trade places on \texttt{LP39}. Proxima's margins over LRC, PropHunt, and AlphaSyndrome remain largely unchanged on \texttt{LP39} and widen on \texttt{HexCC61}---for instance, expanding from $40\times$ to $76\times$ over AlphaSyndrome.
Against ASC, the improvement adjusts from \unsure{$2.6\times$} to $1.4\times$ on \texttt{HexCC61} (with overlapping confidence intervals) and from \unsure{$2.1\times$} to $1.9\times$ on \texttt{LP39}. On \texttt{BB108}, Proxima's depth-8 schedule reaches an LER of $10^{-8}$ compared to $6\times 10^{-8}$ for the depth-7 ASC and reference schedules, though limited failure counts again yield overlapping confidence intervals.

\begin{takeaways}[title={RQ5 Summary}]

    Because proximity captures intrinsic DEM fault structure rather than overfitting to a specific decoder or noise assumption, Proxima's schedules transfer robustly across noise regimes and decoders without re-optimization.
\end{takeaways}

\section{Related Work}
\label{sec:related}

Before recent attention to automated SEC design, schedules were code-specific gadgets. Simple baselines order the parity-check interactions by qubit index, as in early QEC experiments~\cite{ryan-anderson_realization_2021,zhao_realization_2022}, or minimize depth to limit idling~\cite{Constant_overhead}. At a fixed depth, the interaction order decides how correlated faults are oriented. The N- and Z-shaped orderings of the rotated surface code point hook errors away from the logical operators~\cite{fowler_surface_2012,Low_distance}, and a diagonal schedule makes this orientation uniform while shortening the cycle~\cite{kishony_surface_2026}. For the varied layouts of lattice surgery, where plaquette-wise planning breaks down, the diagonal schedule and ZX interleaving of the $Z$ and $X$ tiles keep the full fault distance at no extra depth~\cite{kishony_surface_2026,hirai_no_2026}. The depth-7 SEC schedule of Bravyi et al.~\cite{bravyi_high-threshold_2024} was found by a search over gate orderings that maintain a good circuit-level distance and is the standard baseline for BB code memories. Other primitives target specific families: tangling schedules ease connectivity requirements~\cite{geher_tangling_2024}, morphing circuits lower the qubit degree of BB codes~\cite{shaw_lowering_2025}, and distance-preserving criteria tell when low-depth hypergraph product circuits keep the code distance~\cite{manes_distance-preserving_2025}. Together, these templates fix the design rules a schedule compiler must respect, but each is bound to one code family and carries no optimality guarantee.

Automated schedulers replace such templates by casting SEC design as a compiler optimization problem. One line of work settles the circuit depth. ASC casts the legality and timing constraints of stabilizer measurement as an SMT instance and returns depth-optimal circuits, or near-optimal ones when the solver times out~\cite{zhang_optimal_2026}. LocalASC restricts this search to schedules invariant under a subgroup of the Tanner-graph automorphisms, with one variable per edge orbit, and finds depth-optimal schedules for two-block and quantum Tanner codes on which ASC times out~\cite{durazo_rocha_local_2026}. 
For lifted product families such as BB codes, algebraic constructions reach or come within one layer of the depth lower bound without search~\cite{tan_syndrome_2026,nguyen_disassembling_2026}.
None of these methods further considers the differences in fault-tolerant performance among various parity-check orderings within the chosen depth. Another line of work searches for SEC schedules guided by an estimate of their fault tolerance. PropHunt starts from a coloration circuit and reorders CNOTs to reduce syndrome ambiguity, estimated by MaxSAT on randomly grown DEM subgraphs~\cite{viszlai_prophunt_2026}. AlphaSyndrome explores ordering and parallelism jointly by MCTS with a decoder in the loop, so every rollout pays for sampling and decoding~\cite{liu_alphasyndrome_2026}. FastSched trains a reinforcement-learning policy that picks the CNOT order of each check and scores schedules by an LER estimated with importance sampling, which saves decoded shots~\cite{ye_reinforcement_2026}. LRC generalizes left--right circuits, first proposed for specific code families~\cite{xu_constant-overhead_2024}, to arbitrary CSS codes, staggering $X$- and $Z$-checks without interleaving their gates, and ranks the circuits by residual-error-related metrics~\cite{strikis_high-performance_2026}. These schedulers share the loop of generating candidates and scoring them, but differ in what is scored: depth alone, the ambiguity of sampled DEM fragments, the logical error rate of a decoder, or the residual distance of hook errors.

In the broader context of physical-layer fault-tolerant compilation for syndrome extraction~\cite{zhu2026quantum}, other works address resource overheads orthogonal to pure parity-check ordering. Few-ancilla scheduling reuses ancillas across checks, trading depth against qubit count~\cite{sato_scheduling_2025}, and layout synthesizers map a chosen circuit onto sparse architectures~\cite{yin_qecc-synth_2025}. For the surface code, redesigning the circuit rather than the code relaxes hardware requirements~\cite{mcewen_relaxing_2023} and tolerates fabrication defects without losing spacelike distance~\cite{zhou_routing_2026,debroy_luci_2025}. A separate strand works directly on DEMs, such as the DEM-based design that shapes circuits by their DEM structure~\cite{Derks2025designingfault} and the SAT-based synthesis of state-preparation circuits, which searches for low-weight undetectable fault sets~\cite{peham_automated_2025}.

\section{Conclusion}
\label{sec:conclusion}

We presented {Proxima}, an automated syndrome extraction circuit scheduling framework that bridges structural circuit synthesis and physical fault tolerance for CSS-type qLDPC codes. Rather than relying on bespoke heuristics or costly in-loop simulation and decoding, {Proxima} pairs constraint-based schedule generation with progressive ranking via {proximity}---an analytical surrogate metric on the DEM that captures arbitrary circuit-level fault mechanisms within graph-local neighborhoods. {Proxima} consistently outperforms other SOTA schedulers, reducing logical error rates by 1.4--2.6$\times$ over the strongest baseline, while generalizing robustly across noise models and decoders. By exploiting the translational symmetry of bivariate bicycle codes through a compact monomial representation, {Proxima} identifies uniform depth-8 schedules that cut logical error rates by 2.1--5.1$\times$ compared to IBM's depth-7 reference schedule~\cite{bravyi_high-threshold_2024}. More broadly, these results establish DEM-level fault-propagation analysis as a tractable, high-resolution compiler objective, advancing scalable fault-tolerant compilation toward practical quantum computing.

\begin{acks}
    This work was supported in part by the AI Chip Center for Emerging Smart Systems (ACCESS) under the InnoHK initiative of the Innovation and Technology Commission of the Hong Kong Special Administrative Region Government, and in part by the Research Grants Council of the Hong Kong Special Administrative Region under Grant No.~16217326.
\end{acks}

\bibliographystyle{ACM-Reference-Format}
\bibliography{references}


\end{document}